# Characterizing universal topological phases beyond spatial embedding

Sunkyu Yu[1†], Xianji Piao[2§], and Namkyoo Park[3*]

[1]Intelligent Wave Systems Laboratory, Department of Electrical and Computer Engineering, Seoul National University, Seoul 08826, Korea

[2]Wave Engineering Laboratory, School of Electrical and Computer Engineering, University of Seoul, Seoul 02504, Korea

[3]Photonic Systems Laboratory, Department of Electrical and Computer Engineering, Seoul National University, Seoul 08826, Korea

E-mail address for correspondence: sunkyu.yu@snu.ac.kr; piao@uos.ac.kr; nkpark@snu.ac.kr

**Abstract**

Although Hamiltonians govern wave-matter interactions, conventional characterizations of topological phases of matter require more than the Hamiltonian alone, relying on global invariants defined on closed reciprocal-space manifolds or local commutation relations with position operators. This spatial dependence precludes a unique characterization of topological phases on an abstract platform, whose Hamiltonian can be defined independently of its spatial embedding. Here we establish a framework for characterizing universal topological phases—grounded in the invariance of topological features under arbitrary Hamiltonian-preserving deformations. We construct a Hamiltonian-based spatial embedding that groups a system's constituents according to their connectivity. This embedding, which reproduces the results of conventional topological diagnostics in established systems, enables a coherent extension of topological phases to complex networks using the Hamiltonian alone, further highlighting exceptional robustness in small-world networks. Our results reshape the conceptual foundations of topological physics by decoupling topological characterization from prescribed physical platforms.

## Introduction

Topological phases of matter have become a central theme in modern materials science[1], optics[2], acoustics[3], and quantum physics[4]. A hallmark is the emergence of topologically protected boundary—including end[5], edge[6], and corner[7]—states, which exhibit robust transport or localization under moderate disorder that preserves the relevant gaps and symmetries. Canonical platforms for topological phases across different dimensional spaces[8,9], initially developed for periodic Hermitian systems with short-range hopping, have been extended dramatically to disordered[10-14] and time-varying[15] materials, non-Euclidean geometry[16], nonlocal hopping[17-19], complex networks[20,21], and non-Hermitian systems[22].

Despite extensive generalizations of topological matter, its physical realization inherently relies on embedding constituents within physical domains: real or reciprocal spaces spanned by spatial or temporal axes. Consequently, conventional topological diagnostics—ranging from reciprocal-space invariants like Chern numbers[1] to real-space tools such as Chern markers[23], Bott indices[24], and spectral localizers[25,26]—depend not only on the system's Hamiltonian but also on its physical-domain embedding. This dependence conflates intrinsic topology with the geometry of the physical domain: while constituent-resolved wave dynamics—including those of topologically protected boundary states—remain strictly invariant under any Hamiltonian-preserving transformation[27,28], the topological diagnostics generally vary with the spatial embedding of the constituents. Such embedding-dependent metrics therefore cannot uniquely characterize the topological phases of abstract platforms, such as effective tight-binding models[6], synthetic dimensions with frequency modes[29] or Fock states[30], and complex networks[20,21]. To understand canonical topological matter and these abstract systems through a unified lens, developing an embedding-independent framework compatible with Hamiltonian-preserving rearrangements remains an open challenge.

Here, we propose an embedding-independent characterization of topological phases, applicable to both established topological matter and abstract networks. Using graph Laplacian eigenstates, we construct a Hamiltonian-intrinsic spatial embedding that groups strongly connected sites. This embedding recovers the topological phases of canonical models and their variants without reference to externally assigned site positions. Employing the design procedure in our previous work[21], we investigate this class of topological phases across a broad spectrum of complex networks, revealing their robustness in the small-world regime characterized by high clustering and a shortcut-induced bulk. These results shift the basis of topological characterization to the intrinsic structure of Hamiltonians, providing a natural route to identifying universal topological phases consistent with constituent-resolved wave transport.

## Results

### Embedding dependence of topological diagnostics

In canonical lattices, spatial geometry provides a natural basis for grouping sites according to their proximity. Because spatial proximity generally correlates with the strength of local hopping mediated by scattering or evanescent coupling, the spatial embedding directly reflects the locality of the Hamiltonian. Given this locality, conventional topological diagnostics[1,23,24,26] provide well-established characterizations of topological phases.

However, this geometric basis is not uniquely defined in abstract systems lacking an intrinsic spatial embedding, such as synthetic dimensions that combine different physical axes[29-31], structures projected between geometries with different curvatures[16], and complex networks described by tight-binding models[20,21]. To evaluate the effect of this embedding ambiguity, we use a representative topological diagnostic—spectral localizer[25,26]—that is connected to other diagnostics for periodic or aperiodic systems[32-34]. In a two-dimensional (2D) system, the spectral localizer $L$ evaluated at energy $E$ and at the $n$-th site is

$$L = \left(H - EI\right)\otimes\sigma_z + \kappa\left[\left(X - x_n I\right)\otimes\sigma_x + \left(Y - y_n I\right)\otimes\sigma_y\right], \tag{1}$$

where $(x_n, y_n)$ denotes the position of the $n$-th site, $X$ and $Y$ are the position operators determined by the site coordinates, $\sigma_{x,y,z}$ denote the Pauli matrices, $\kappa$ is a hyperparameter that balances the spatial and energetic contributions, and $I$ is the identity matrix. Using this expanded operator, we can obtain two critical metrics: $C_\mathrm{L}$ = sig($L$)/2—the local Chern index that quantifies position-energy local topology—and $\mu_\mathrm{L}$ = min[|Spec($L$)|]—the local gap that quantifies the robustness of $C_\mathrm{L}$—where sig($L$) and Spec($L$) denote the signature and spectrum of $L$, respectively.

Figure 1 illustrates how these diagnostics strongly depend on the spatial embedding. We consider the conventional Hofstadter model on a finite square lattice (see Methods). In the canonical embedding, $C_\mathrm{L}$ possesses the nonzero quantized value of +1 throughout the bulk and vanishes at the boundary (Fig. 1a), which yields a chiral edge state (Fig. 1b) as expected from the bulk-edge correspondence[26]. While preserving the original Hamiltonian of this model, we apply an inside-out embedding to the canonical lattice, which relocates the edge sites to the interior and scatters the bulk sites around them (Figs. 1c,1d). Because the eigenstates are determined solely by the Hamiltonian, the edge state is preserved under this transformation (Fig. 1d). In contrast, from the localizer, which correlates the Hamiltonian with the transformed position operators $X$ and $Y$, the local Chern index collapses to $C_\mathrm{L}$ = 0 everywhere (Fig. 1c). This result demonstrates that the conventional diagnostics no longer identify the topological phase that evidently persists against the transformation.

The same phenomenon appears in topological insulating phases of complex networks (Figs. 1e-1h). For a small-world topological insulator designed under the same procedure in our previous work[21] (see Methods), the embedding optimized throughout the design process yields a $C_L = +1$ at the bulk surrounded by a trivial edge (Fig. 1e). This nontrivial local topology leads to a directional state analogous to a chiral edge state (Fig. 1f), which is robust against disorder[21]. However, replacing this embedding by uniformly random coordinates erases the topological signature (Fig. 1g) despite the conservation of a directional state (Fig. 1h).

Figure 1 demonstrates that the conventional spectral localizer fails to uniquely characterize topological phases when the spatial embedding is not fixed by the Hamiltonian. Similar limitations arise in other conventional diagnostics. For example, the Bloch-band Chern number requires a Brillouin zone associated with discrete translational symmetry, which is not generally available for an abstract network. Real-space diagnostics, such as Chern markers[23] and Bott indices[24], combine the spectral projector of $H$ with the position operators $X$ and $Y$, enforcing the dependence on the chosen spatial embedding. In exploring topological phases in abstract platforms, these limitations motivate an alternative diagnostic that relies solely on the Hamiltonian, while consistently reproducing the results of conventional diagnostics for canonical models.

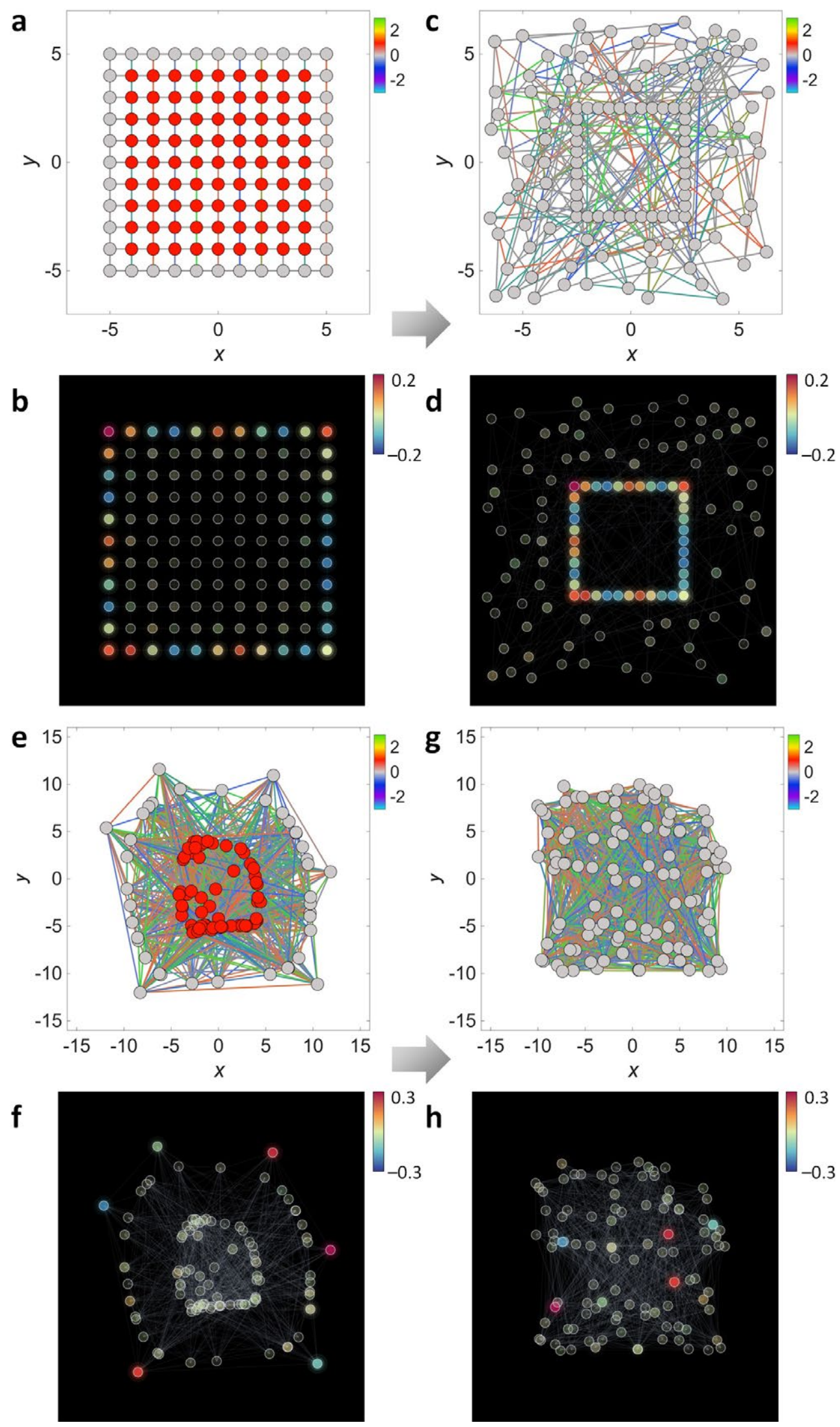

**Fig. 1. Conventional diagnostics cannot uniquely probe the topological phases of abstract systems. a-d**, Hofstadter model on a finite square lattice. **a**, $C_L$ on the canonical embedding. **b**, Real part of the chiral edge state. **c,d**, The same quantities after an inside-out embedding. **e-h**, Small-world topological insulator. **e,f**, On an embedding that separates bulk and boundary nodes, $C_L = +1$ (**e**) and the directional eigenstate (**f**) are obtained. **g,h**, On random coordinates, the same Hamiltonian gives a trivial diagnostic (**g**) despite the conservation of the directional state (**h**).

## Hamiltonian-derived embedding

To overcome the embedding dependence of topological diagnostics, we introduce an embedding derived from the Hamiltonian itself for evaluating the spectral localizer (Figs. 2a-2c). This embedding is designed to reproduce the results of conventional diagnostics by grouping the sites that are strongly coupled in the Hamiltonian, analogous to the way canonical lattices group neighbouring sites in space.

To develop this analogy, the connectivity structure of the Hamiltonian is encoded in an adjacency matrix $A$ with $A_{mn} = |H_{mn}|$, which yields the corresponding graph Laplacian $G = D - A$ with the degree matrix $D$ (Fig. 2a). Because $G$ is real symmetric and positive semidefinite, the eigenvalue problem $Gv_m = \lambda_m v_m$ ($m = 0, 1, \ldots, N - 1$) produces an orthonormal eigenbasis[35] ordered by increasing eigenvalue, with $\lambda_0 = 0$ and the constant vector $v_0 \propto (1, \ldots, 1)^{\mathrm{T}}$. Because $v_m^{\mathrm{T}} G v_m = \sum_{i<j} A_{ij}[v_m(i) - v_m(j)]^2 = \lambda_m$, the lowest-order eigenstates with small $\lambda_m$ group strongly coupled sites: $|v_m(i) - v_m(j)|$ is small wherever $A_{ij}$ is large (see Methods).

Using these properties of the Laplacian eigenstates, we construct the Hamiltonian-derived embedding. To diagnose $d$-dimensional topological features, we compose the coordinates $(x_1, x_2, \ldots, x_d)_i$ of the $i$-th site from the $d$ lowest-order nontrivial eigenstates that are locally independent of one another, ensuring that the coordinates vary along locally distinct directions (see Methods and Supplementary Note S1). In the Hofstadter example, this configuration yields $(x, y)_i = [v_1(i), v_2(i)]$ (Fig. 2b), resulting in an embedding that recovers the square geometry up to a rotation and slight warping near the edges, while depending solely on $H$ (Fig. 2c).

We note that this Hamiltonian-derived embedding with $X = X(H)$ and $Y = Y(H)$ successfully reproduces the topological characteristics of the Hofstadter lattice without any use of the prescribed site positions (Supplementary Note S2). The quantized index $C_{\mathrm{L}} = +1$, together with a finite local gap $\mu_{\mathrm{L}}$, is obtained throughout the original bulk sites, whereas $C_{\mathrm{L}} = 0$ at the original edge sites (Fig. 2d). Scanning the flux $\alpha$ per plaquette and the energy $E$ also reproduces the familiar Hofstadter butterfly for this finite-size network (Fig. 2e).

To demonstrate the generality of this embedding, we apply the same procedure to other canonical models and their variants (see Methods): topological phases in a kagome lattice (Fig. 2f), a hyperbolic lattice (Fig. 2g), a stealthy hyperuniform (SHU) lattice (Fig. 2h), and the Haldane model (Fig. 2i). In all cases, the Hamiltonian-intrinsic embedding preserves the topological characteristics of the original models (Supplementary Note S3) without reference to any external geometry. This observation demonstrates that a prescribed spatial arrangement of sites is not essential at least for characterizing topological phases conventionally defined in systems with local Hamiltonians.

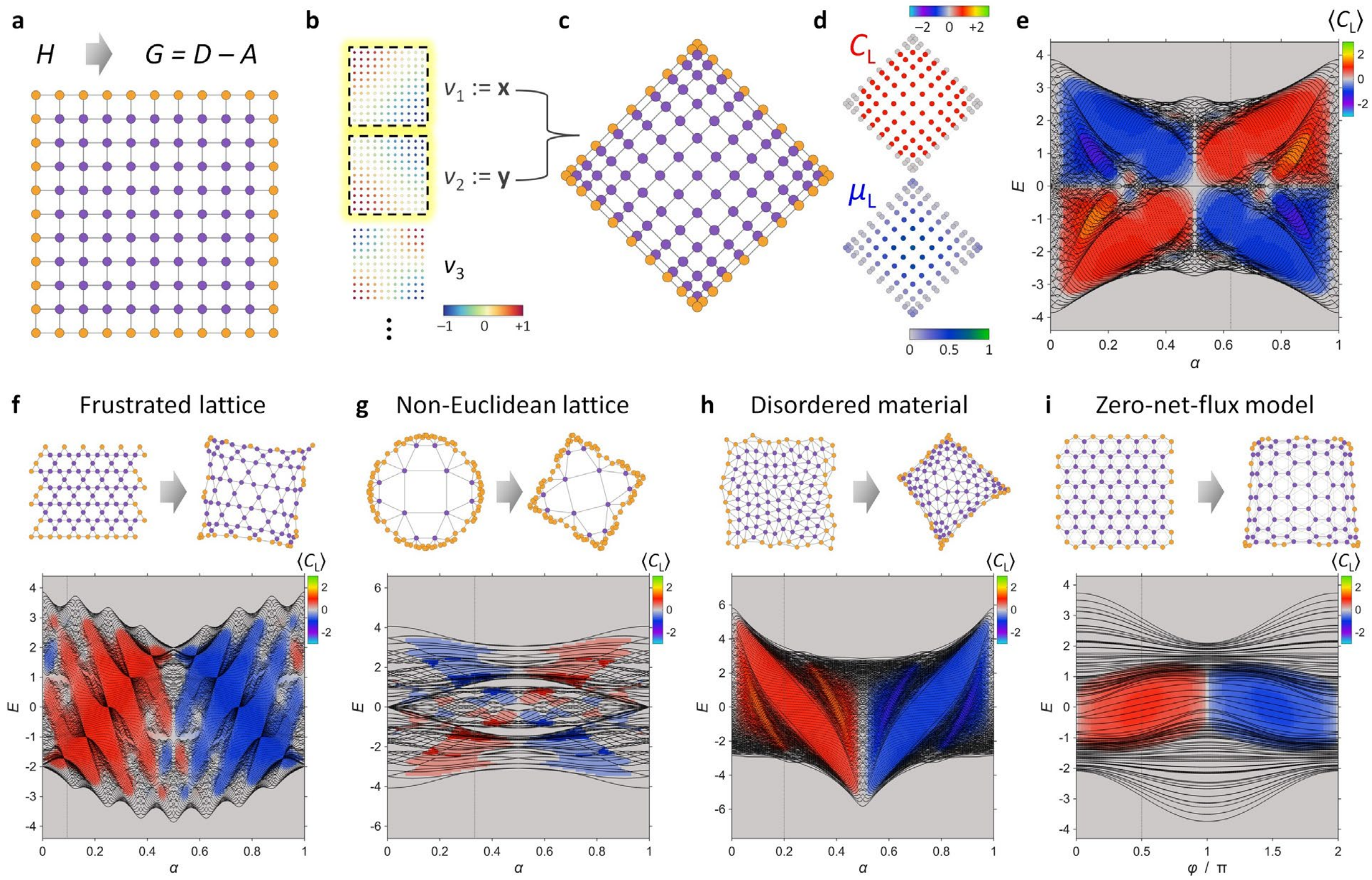

**Fig. 2. Topological phases beyond spatial embedding. a,** The graph Laplacian $G$ is built from $H$. **b**, The lowest nontrivial eigenstates $v_1$, $v_2$, … of $G$, where $v_1$ and $v_2$ define the site coordinates **x** and **y**. **c,d**, The Hamiltonian-derived embedding of the Hofstadter lattice (**c**) and the resulting $C_L$ and $\mu_L$ (**d**). **e**, Bulk-averaged $\langle C_L \rangle$ on the Hamiltonian-derived embedding as a function of flux per plaquette $\alpha$ and energy $E$, with the eigenvalue spectrum (black dots). **f-i**, Reference and Hamiltonian-derived embeddings (top) and the corresponding butterflies (bottom) for a kagome lattice (**f**), a hyperbolic {4, 6} lattice (**g**), a disordered SHU (**h**), and the Haldane model as a function of the next-nearest-neighbour phase $\varphi$ (**i**). $\alpha = +\Phi/2\pi$ in **e** and **g** and $\alpha = -\Phi/2\pi$ in **f** and **h**, where $\Phi$ is the flux through one plaquette. Purple and amber sites in **a**, **c**, and **f-i** denote the bulk and edge of the original lattices, respectively. Vertical lines in **e-i** denote the design points for $C_L$ and $\mu_L$ in **d** and Supplementary Note S3.

## Network topological phases

Because the Hamiltonian-derived embedding successfully reproduces the topological phases of local Hamiltonians, we extend this embedding to nonlocal complex networks to explore topological phases that correspond to a coherent generalization of local ones. We investigate which network architectures can support topological phases by examining Watts-Strogatz networks[36], in which each link of a ring lattice is rewired with probability $p$, spanning regular, small-world, and random regimes (Supplementary Note S4). For every network, we fix the connectivity, derive the embedding from $G$, and optimize only the link gauge fields following the procedure of our previous work[21]. The procedure aims to ensure that the innermost sites (60 % or 30 %) of a network acquire $C_L = +1$, while its outer sites (40 % or 70 %) remain trivial at $E = 1$.

Figures 3a-3c illustrate the Hamiltonian-derived embedding (left) and the resulting topological diagnostics (middle: $C_L$, right: $\mu_L$) across different network topologies. For the regular lattice (Fig. 3a), the two lowest Laplacian eigenvectors used in the embedding correspond to the spatial harmonics of the ring; the derived embedding is thus a hollow circle where all nodes are geometrically equivalent. This uniform radial positioning precludes any bulk-edge separation, yielding $C_L = 0$ throughout the network regardless of the link gauge fields—signifying the absence of 2D topological phases. In contrast, modest rewiring into a small-world network (Fig. 3b with $p = 0.1$) readily lifts this positional degeneracy: shortcut links populate the interior of the embedding, starting to draw bulk sites inward. This manifestation of clear bulk-edge separation enables the realization of a $C_L = +1$ core enclosed by a trivial edge, accompanied by a robust, finite local gap within the nontrivial bulk. For the random network (Fig. 3c), the embedding collapses into a dense central core with peripheral low-degree nodes. Although a nontrivial bulk is still successfully formed, the local gap diminishes substantially across the bulk, and the largest values of $\mu_L$ occur only at isolated peripheral nodes, rendering the resulting topological phase fragile.

To verify this picture, we scan over $p$ for different target bulk fractions (Fig. 3d: 60 % and 30 % bulk sites). Consistent with the behaviour shown in Figs. 3a-3c, the bulk-averaged local

Chern index rises from zero at $p = 0$ to $\langle C_L \rangle \approx 0.9$ at $p = 0.1$, saturating near unity for $p \geq 0.1$. In contrast, the bulk local gap exhibits a non-monotonic dependence on $p$: it peaks at $\langle \mu_L \rangle \approx 0.2$ for $p = 0.1$ to 0.3 and drops to $\langle \mu_L \rangle \approx 0.1$ at $p = 1$. The optimum thus lies at the upper edge of the small-world regime, where the network contains enough shortcuts for the embedding to separate the bulk from the edge, yet retains sufficient residual clustering to preserve a robust local gap. Conversely, regular networks fail because their embeddings cannot define a bulk, whereas random networks support a nontrivial topology only marginally due to the fragility of the resulting topological phase. These findings highlight the critical impact of underlying network architecture on wave topology—spatial bulk characterization for nontrivial topological phases and structural clustering for their robustness—simultaneously underscoring the clear advantage of the small-world architecture.

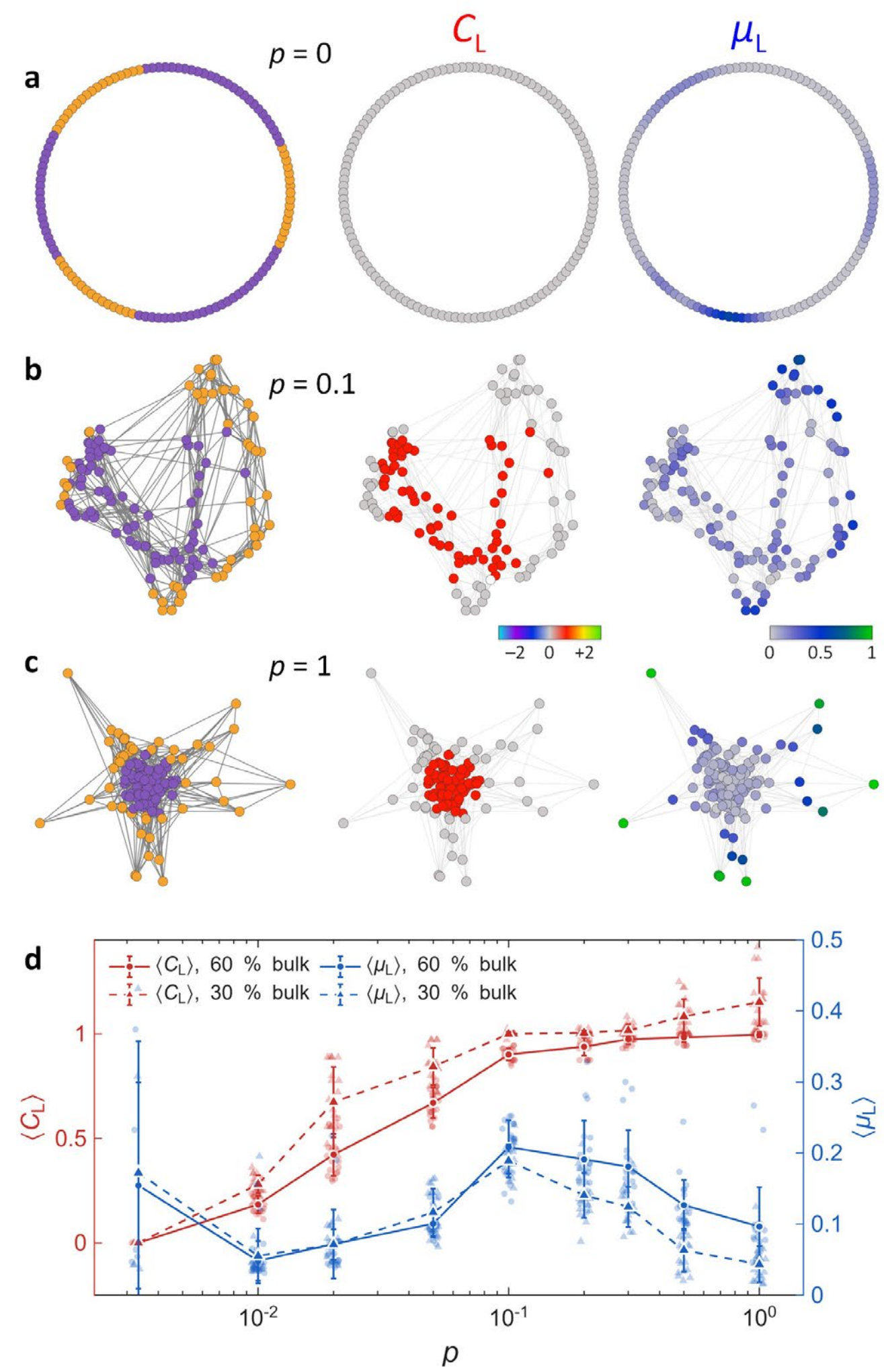

**Fig. 3. Topological phases of Watts-Strogatz networks. a-c**, Embedding (left; bulk, purple; edge, amber), $C_L$ (middle) and $\mu_L$ (right) after optimization[21], for a regular lattice, $p = 0$ (**a**), a small-world network, $p = 0.1$ (**b**), and a random network, $p = 1$ (**c**). The number of sites is $N = 120$ and the mean node degree is $k = 10$. In **a-c**, the networks are optimized to achieve $C_L = +1$ at the bulk: the innermost 60 % of the sites. In **a**, the bulk and edge sites are determined only by numerical round-off in the radial ordering. **d**, Bulk-averaged $\langle C_L \rangle$ (red) and $\langle \mu_L \rangle$ (blue) at the design point versus $p$. Transparent symbols mark the 20 realizations per $p$ (four network realizations with five optimization seeds). Filled symbols and error bars denote the mean and standard deviation. The bulk is defined as the innermost 60 % or 30 % of the embedding. Reducing the fraction of nontrivial sites to 30 % allows higher $\langle C_L \rangle$, yet the smaller bulk core weakens local gaps. The leftmost points represent $p = 0$.

## Discussion

Although the Hamiltonian-derived embedding identifies topological phases that are conserved under any change of embedding, the embedding used here—the lowest-order eigenvectors of the graph Laplacian—is only one of the constructions that generalize the conventional diagnostics of canonical models to abstract platforms including nonlocal networks; it does not capture every topological phase a complex network can host. This limitation is evident in the topological phases designed in our previous work[21], where the link gauge fields and the site positions were optimized jointly: the designed local Chern indices are not recovered using our Laplacian embedding (Supplementary Note S5). In this context, whether a universal embedding exists that captures all such topological phases remains an open question. As a step toward addressing this remaining challenge, other constructions can be explored that preserve the same correspondence with canonical models while grouping sites by different notions of proximity: the normalized graph Laplacian[35], which compensates for strongly varying onsite potentials and degrees; diffusion maps[37], in which the distance between sites is the diffusion distance; and the magnetic graph Laplacian[30,38], whose eigenstates retain the hopping phases and thus encode the gauge-field frustration. Because each of these preserves the locality of the Hamiltonian, each extends conventional topological phases to complex networks in different contexts.

In conclusion, we have developed a method for revealing topological phases that are universal with respect to the spatial embedding—that is, preserved under any change of the site coordinates. For 2D examples, we devised a spatial embedding determined solely by the Hamiltonian, which characterizes the topological phases intrinsic to the Hamiltonian and reproduces the conventional topological phases of various canonical models. We extended these embedding-universal topological phases to complex networks, demonstrating the advantage of small-world networks, in which shortcuts yield a bulk-edge separation and clustering provides robustness. Although complex networks can host a far richer variety of topological phases than those probed by our methodology, our approach paves the way for separating physical-space information from topological phases in matter and networks. The method also extends to topological phenomena in other dimensions, such as topologically protected end and corner states.

## Methods

**Hofstadter model.** We use the square-lattice Hofstadter model of 11 × 11 sites with nearest-neighbour hopping $t = 1$ and a flux of 1.25π per plaquette in the Landau gauge in Figs. 1a and 1b. The localizer is evaluated at $E = 1$ with $\kappa = 0.25$.

**Small-world topological insulator.** The network of Figs. 1e and 1f is a Watts-Strogatz network[36] ($N = 100$, degree 20 and rewiring probability $10^{-3}$) whose link gauge fields and site positions were jointly optimized using the method in our previous work[21] for $C_L = +1$ on the inner 60 % of the sites at $E = 1$ and $\kappa = 0.25$. The directional state is the eigenstate nearest $E$.

**Graph Laplacian eigenstates.** Consider the $N$-dimensional graph Laplacian $G = D - A$, whose elements are $[G]_{ij} = \delta_{ij}\sum_k A_{ik} - A_{ij}$, where $\delta_{ij}$ is the Kronecker delta function. For an arbitrary $N \times 1$ state $u$, we obtain

$$\begin{aligned} u^{\mathrm{T}} G u &= \sum_{i,j} u(i) \left( \delta_{ij} \sum_k A_{ik} - A_{ij} \right) u(j) \\ &= \sum_i \sum_k A_{ik} u^2(i) - \sum_{i,j} A_{ij} u(i) u(j) \\ &= \sum_{i<j} A_{ij} [u(i) - u(j)]^2. \end{aligned} \tag{2}$$

where $u(i)$ denotes the $i$-th element of $u$. Because $G v_m = \lambda_m v_m$ for the normalized eigenstates[35], we obtain

$$\lambda_m = \sum_{i<j} A_{ij}[v_m(i) - v_m(j)]^2, \tag{3}$$

which leads to the bound:

$$|v_m(i) - v_m(j)| \leq \sqrt{\lambda_m / A_{ij}}. \tag{4}$$

For the lowest-order eigenstates with small $\lambda_m$, Eq. (4) tightly bounds the separation between the $i$-th and $j$-th sites when $A_{ij}$ is large. Therefore, the embedding using the graph Laplacian eigenstates can provide the grouping of strongly coupled sites according to the adjacency matrix $A$.

**Selection of locally independent eigenstates.** In our $d$-dimensional embedding, the first axis is $x_1 = v_1$, and the remaining $d - 1$ axes are selected from the lowest-order eigenstates that are locally independent of those already chosen[39]. For $d = 2$ with $x = v_1$, the candidates for $y = x_2$ are $\{v_b\}$ ($b = 2, 3, \ldots, 5$). Each $v_b$ is scored by the local nondegeneracy of the map ($v_1$, $v_b$) with the link difference vectors $\delta_{ij} = [v_1(j) - v_1(i), v_b(j) - v_b(i)]^{\mathrm{T}}$ over the neighbours $j$ of site $i$. Using $\delta_{ij}$, the weighted Gram matrix $g_i = \sum_j A_{ij}\delta_{ij}\delta_{ij}^{\mathrm{T}}$ represents the local dispersion of the neighbours of site $i$. We calculate the normalized measure $\eta_i = 4\det(g_i)/(\mathrm{Tr}[g_i])^2 \in [0, 1]$ from $g_i$, which vanishes where the map folds onto a line and reaches unity where it is locally isotropic. The candidate maximizing the minimum of $\eta_i$ over the sites with at least three links, that is, the eigenstate most locally independent of $v_1$, is selected as $y$.

**Kagome lattices.** We use a frustrated kagome lattice of 6 × 6 unit cells with nearest-neighbour hopping $t = 1$ under a uniform flux of $-1.5\pi$ per unit cell, distributed over its triangular and hexagonal plaquettes in proportion to their areas[40]. The localizer is evaluated at $E = -1.2$ with $\kappa = 0.25$.

**Hyperbolic lattices.** The {4, 6} hyperbolic lattice is generated to three generations in the Poincaré disk. Each face carries the same flux of $2\pi/3$, assigned face by face through an evolving gauge assignment algorithm[16]. The localizer is evaluated at $E = -1.5$ with $\kappa = 0.25$.

**Disordered SHU.** With the collective-coordinate method[41], $N$ = 121 points are generated by minimizing the structure factor within $|k| < K = 4.91$ for $\chi = 0.5$ to satisfy the SHU condition[42,43], where $\chi$ is the dimensionless ratio of constrained to total degrees of freedom. The network links are obtained by the Delaunay triangulation, while the links longer than 1.75 times the mean spacing are removed. Sites lying within one spacing of the boundary are set to be edge sites. Each Delaunay face carries a flux of $-0.4\pi$. The spectral localizer is evaluated at $E = 1.1$.

**Haldane model.** The Haldane model on a 7 × 9 honeycomb is developed with nearest-neighbour hopping $t_1 = 1$, next-nearest-neighbour hopping $t_2 = 0.15 \times \exp(\pm i\varphi)$, and zero sublattice mass $M = 0$, satisfying the condition that the net flux per unit cell vanishes[44]. The design point is $\varphi = \pi/2$ at $E = 0$ with $\kappa = 0.25$. The butterfly in Fig. 2i scans $\varphi/\pi$ over [0, 2].

## Data availability

Data used in the current study are available from the corresponding authors upon request and can also be obtained by running the shared codes at [DOI] in the Zenodo.

## Code availability

All original code has been deposited in Zenodo and is publicly available at [DOI].

## Acknowledgements

We acknowledge financial support from the National Research Foundation of Korea (NRF) through the Core Research Grants (no. RS-2026-25469085), Innovation Research Center (no. RS-

2024-00413957), Young Researcher Program (no. RS-2025-00552989), Pilot and Feasibility Grants (no. RS-2025-19912971), and Midcareer Researcher Program (no. RS-2023-00274348), all funded by the Korean government. This work was supported by the BK21 FOUR program of the Education and Research Program for Future ICT Pioneers in 2026, through Seoul National University. We also acknowledge administrative support from SOFT foundry institute.

## Author Contributions

S.Y., X.P., and N.P. contributed equally to conceiving the idea, developing the code, discussing the results, and preparing the final manuscript.

## Competing Interests

The authors have no conflicts of interest to declare.

## Additional information

**Correspondence and requests for materials** should be addressed to S.Y., X.P., or N.P.

## Figure Legends

**Fig. 1. Conventional diagnostics cannot uniquely probe the topological phases of abstract systems. a-d**, Hofstadter model on a finite square lattice. **a**, $C_L$ on the canonical embedding. **b**, Real part of the chiral edge state. **c,d**, The same quantities after an inside-out embedding. **e-h**, Small-world topological insulator. **e,f**, On an embedding that separates bulk and boundary nodes, $C_L = +1$ (**e**) and the directional eigenstate (**f**) are obtained. **g,h**, On random coordinates, the same Hamiltonian gives a trivial diagnostic (**g**) despite the conservation of the directional state (**h**).

**Fig. 2. Topological phases beyond spatial embedding. a,** The graph Laplacian $G$ is built from $H$. **b**, The lowest nontrivial eigenstates $v_1$, $v_2$, … of $G$, where $v_1$ and $v_2$ define the site coordinates **x** and **y**. **c,d**, The Hamiltonian-derived embedding of the Hofstadter lattice (**c**) and the resulting $C_L$ and $\mu_L$ (**d**). **e**, Bulk-averaged $\langle C_L \rangle$ on the Hamiltonian-derived embedding as a function of flux per plaquette $\alpha$ and energy $E$, with the eigenvalue spectrum (black dots). **f-i**, Reference and Hamiltonian-derived embeddings (top) and the corresponding butterflies (bottom) for a kagome lattice (**f**), a hyperbolic {4, 6} lattice (**g**), a disordered SHU (**h**), and the Haldane model as a function of the next-nearest-neighbour phase $\varphi$ (**i**). $\alpha = +\Phi/2\pi$ in **e** and **g** and $\alpha = -\Phi/2\pi$ in **f** and **h**, where $\Phi$ is the flux through one plaquette. Purple and amber sites in **a**, **c**, and **f-i** denote the bulk and edge of the original lattices, respectively. Vertical lines in **e-i** denote the design points for $C_L$ and $\mu_L$ in **d** and Supplementary Note S3.

**Fig. 3. Topological phases of Watts-Strogatz networks. a-c**, Embedding (left; bulk, purple; edge, amber), $C_L$ (middle) and $\mu_L$ (right) after optimization[21], for a regular lattice, $p = 0$ (**a**), a small-world network, $p = 0.1$ (**b**), and a random network, $p = 1$ (**c**). The number of sites is $N = 120$ and the mean node degree is $k = 10$. In **a-c**, the networks are optimized to achieve $C_L = +1$ at the bulk: the innermost 60 % of the sites. In **a**, the bulk and edge sites are determined only by numerical round-off in the radial ordering. **d**, Bulk-averaged $\langle C_L \rangle$ (red) and $\langle \mu_L \rangle$ (blue) at the design point versus $p$. Transparent symbols mark the 20 realizations per $p$ (four network realizations with five optimization seeds). Filled symbols and error bars denote the mean and standard deviation. The bulk is defined as the innermost 60 % or 30 % of the embedding. Reducing the fraction of nontrivial sites to 30 % allows higher $\langle C_L \rangle$, yet the smaller bulk core weakens local gaps. The leftmost points represent $p = 0$.

## References

1. Hasan, M. Z. & Kane, C. L. Colloquium: Topological insulators. *Rev. Mod. Phys.* **82**, 3045-3067 (2010).
2. Ozawa, T. *et al.* Topological photonics. *Rev. Mod. Phys.* **91**, 015006 (2019).
3. Xue, H., Yang, Y. & Zhang, B. Topological acoustics. *Nat. Rev. Mater.* **7**, 974-990 (2022).
4. Nayak, C., Simon, S. H., Stern, A., Freedman, M. & Sarma, S. D. Non-Abelian anyons and topological quantum computation. *Rev. Mod. Phys.* **80**, 1083-1159 (2008).
5. Gröning, O., Wang, S., Yao, X., Pignedoli, C. A., Borin Barin, G., Daniels, C., Cupo, A., Meunier, V., Feng, X. & Narita, A. Engineering of robust topological quantum phases in graphene nanoribbons. *Nature* **560**, 209-213 (2018).
6. Hafezi, M., Demler, E. A., Lukin, M. D. & Taylor, J. M. Robust optical delay lines with topological protection. *Nat. Phys.* **7**, 907-912 (2011).
7. Xie, B.-Y., Wang, H.-F., Wang, H.-X., Zhu, X.-Y., Jiang, J.-H., Lu, M.-H. & Chen, Y.-F. Second-order photonic topological insulator with corner states. *Phys. Rev. B* **98**, 205147 (2018).
8. Schnyder, A. P., Ryu, S., Furusaki, A. & Ludwig, A. W. W. Classification of topological insulators and superconductors in three spatial dimensions. *Phys. Rev. B* **78**, 195125, doi:10.1103/PhysRevB.78.195125 (2008).
9. Chiu, C.-K., Teo, J. C., Schnyder, A. P. & Ryu, S. Classification of topological quantum matter with symmetries. *Rev. Mod. Phys.* **88**, 035005 (2016).
10. Li, J., Chu, R.-L., Jain, J. K. & Shen, S.-Q. Topological anderson insulator. *Phys. Rev. Lett.* **102**, 136806 (2009).
11. Stützer, S., Plotnik, Y., Lumer, Y., Titum, P., Lindner, N. H., Segev, M., Rechtsman, M. C.

& Szameit, A. Photonic topological Anderson insulators. *Nature* **560**, 461-465 (2018).

12. Agarwala, A. & Shenoy, V. B. Topological Insulators in Amorphous Systems. *Phys. Rev. Lett.* **118**, 236402, doi:10.1103/PhysRevLett.118.236402 (2017).

13. Mitchell, N. P., Nash, L. M., Hexner, D., Turner, A. M. & Irvine, W. T. Amorphous topological insulators constructed from random point sets. *Nat. Phys.* **14**, 380-385 (2018).

14. Zhang, Z., Delplace, P. & Fleury, R. Anomalous topological waves in strongly amorphous scattering networks. *Sci. Adv.* **9**, eadg3186 (2023).

15. Galiffi, E., Tirole, R., Yin, S., Li, H., Vezzoli, S., Huidobro, P. A., Silveirinha, M. G., Sapienza, R., Alù, A. & Pendry, J. B. Photonics of time-varying media. *Adv. Photonics* **4**, 014002-014002 (2022).

16. Yu, S., Piao, X. & Park, N. Topological hyperbolic lattices. *Phys. Rev. Lett.* **125**, 053901, doi:10.1103/PhysRevLett.125.053901 (2020).

17. Sticlet, D. & Piéchon, F. Distant-neighbor hopping in graphene and Haldane models. *Phys. Rev. B* **87**, 115402 (2013).

18. Kim, G., Suh, J., Lee, D., Park, N. & Yu, S. Long-range-interacting topological photonic lattices breaking channel-bandwidth limit. *Light Sci. Appl.* **13**, 189, doi:10.1038/s41377-024-01557-4 (2024).

19. Li, M., Zhirihin, D., Gorlach, M., Ni, X., Filonov, D., Slobozhanyuk, A., Alu, A. & Khanikaev, A. B. Higher-order topological states in photonic kagome crystals with long-range interactions. *Nat. Photon.* **14**, 89-94 (2020).

20. Yu, S., Piao, X. & Park, N. Topological heavy-tailed networks. *Newton*, doi:10.1016/j.newton.2026.100598.

21. Yu, S., Piao, X. & Park, N. Topological insulators on complex networks. *arXiv preprint*

*arXiv:2610.02888* (2026).

22. Bergholtz, E. J., Budich, J. C. & Kunst, F. K. Exceptional topology of non-Hermitian systems. *Rev. Mod. Phys.* **93**, 015005 (2021).

23. Bianco, R. & Resta, R. Mapping topological order in coordinate space. *Phys. Rev. B* **84**, 241106 (2011).

24. Loring, T. A. & Hastings, M. B. Disordered topological insulators via C*-algebras. *EPL (Europhysics Letters)* **92**, 67004 (2010).

25. Loring, T. A. K-theory and pseudospectra for topological insulators. *Annals of Physics* **356**, 383-416 (2015).

26. Cerjan, A. & Loring, T. A. Classifying photonic topology using the spectral localizer and numerical K-theory. *APL Photonics* **9**, 111102, doi:10.1063/5.0239018 (2024).

27. Kollár, A. J., Fitzpatrick, M. & Houck, A. A. Hyperbolic lattices in circuit quantum electrodynamics. *Nature* **571**, 45-50, doi:10.1038/s41586-019-1348-3 (2019).

28. Yu, S., Piao, X., Hong, J. & Park, N. Interdimensional optical isospectrality inspired by graph networks. *Optica* **3**, 836-839 (2016).

29. Yuan, L., Lin, Q., Xiao, M. & Fan, S. Synthetic dimension in photonics. *Optica* **5**, 1396-1405 (2018).

30. Yu, S., Piao, X. & Park, N. Magnetic graphs for cavity quantum electrodynamics. *Sci. Adv.* **12**, eaee5566 (2026).

31. Lustig, E., Weimann, S., Plotnik, Y., Lumer, Y., Bandres, M. A., Szameit, A. & Segev, M. Photonic topological insulator in synthetic dimensions. *Nature* **567**, 356-360, doi:10.1038/s41586-019-0943-7 (2019).

32. Jezequel, L., Bardarson, J. H. & Grushin, A. G. Explicit equivalence between the spectral

localizer and local Chern and winding markers. *arXiv preprint arXiv:2508.00214* (2025).

33. Loring, T. A. A guide to the Bott index and localizer index. *arXiv preprint arXiv:1907.11791* (2019).

34. Lozano Viesca, E., Schober, J. & Schulz-Baldes, H. Chern numbers as half-signature of the spectral localizer. *J. Math. Phys.* **60**, 072101 (2019).

35. Chung, F. R. *Spectral graph theory* (American Mathematical Soc., 1997).

36. Watts, D. J. & Strogatz, S. H. Collective dynamics of 'small-world' networks. *Nature* **393**, 440-442, doi:10.1038/30918 (1998).

37. Coifman, R. R., Lafon, S., Lee, A. B., Maggioni, M., Nadler, B., Warner, F. & Zucker, S. W. Geometric diffusions as a tool for harmonic analysis and structure definition of data: Diffusion maps. *Proc. Natl. Acad. Sci. USA* **102**, 7426-7431 (2005).

38. Lange, C., Liu, S., Peyerimhoff, N. & Post, O. Frustration index and Cheeger inequalities for discrete and continuous magnetic Laplacians. *Calc. Var. Partial Differential Equations* **54**, 4165-4196, doi:10.1007/s00526-015-0935-x (2015).

39. Dsilva, C. J., Talmon, R., Coifman, R. R. & Kevrekidis, I. G. Parsimonious representation of nonlinear dynamical systems through manifold learning: A chemotaxis case study. *Applied and Computational Harmonic Analysis* **44**, 759-773 (2018).

40. Ohgushi, K., Murakami, S. & Nagaosa, N. Spin anisotropy and quantum Hall effect in the kagomé lattice: Chiral spin state based on a ferromagnet. *Phys. Rev. B* **62**, R6065 (2000).

41. Uche, O. U., Stillinger, F. H. & Torquato, S. Constraints on collective density variables: Two dimensions. *Phys. Rev. E* **70**, 046122 (2004).

42. Torquato, S. & Stillinger, F. H. Local density fluctuations, hyperuniformity, and order metrics. *Phys. Rev. E* **68**, 041113 (2003).

43. Batten, R. D., Stillinger, F. H. & Torquato, S. Classical disordered ground states: Super-ideal gases and stealth and equi-luminous materials. *J. Appl. Phys.* **104**, 033504 (2008).

44. Haldane, F. D. M. Model for a quantum Hall effect without Landau levels: Condensed-matter realization of the" parity anomaly". *Phys. Rev. Lett.* **61**, 2015 (1988).

# Supplementary Information for "Characterizing universal topological phases beyond spatial embedding"

Sunkyu Yu[1*], Xianji Piao[2†], and Namkyoo Park[3§]

[1]Intelligent Wave Systems Laboratory, Department of Electrical and Computer Engineering, Seoul National University, Seoul 08826, Korea

[2]Wave Engineering Laboratory, School of Electrical and Computer Engineering, University of Seoul, Seoul 02504, Korea

[3]Photonic Systems Laboratory, Department of Electrical and Computer Engineering, Seoul National University, Seoul 08826, Korea

E-mail address for correspondence: [*]sunkyu.yu@snu.ac.kr, [†]piao@uos.ac.kr, [§]nkpark@snu.ac.kr

## Supplementary Note S1. Rectangular geometry of the Hofstadter lattice

In the main text, the two lowest nontrivial eigenstates $v_1$ and $v_2$ serve as the coordinates. This choice is not always suitable in terms of achieving local independence. As an example, we examine a rectangular Hofstadter lattice with the same flux per plaquette as in Figs. 2a-2e of the main text (Fig. S1a). In this geometry, $v_2$ is the second harmonic along the long axis, failing to yield locally distinct directions with $v_1$—folding the lattice onto a one-dimensional curve. Our selection rule described in Methods addresses this ambiguity by employing ($v_1$, $v_3$) as coordinates (Fig. S1b). The resulting embedding reproduces the rectangular lattice up to a smooth distortion (Fig. S1c), and the spectral localizer recovers $C_L$ and $\mu_L$ (Fig. S1d), in agreement with the result on the original embedding (Fig. S1e).

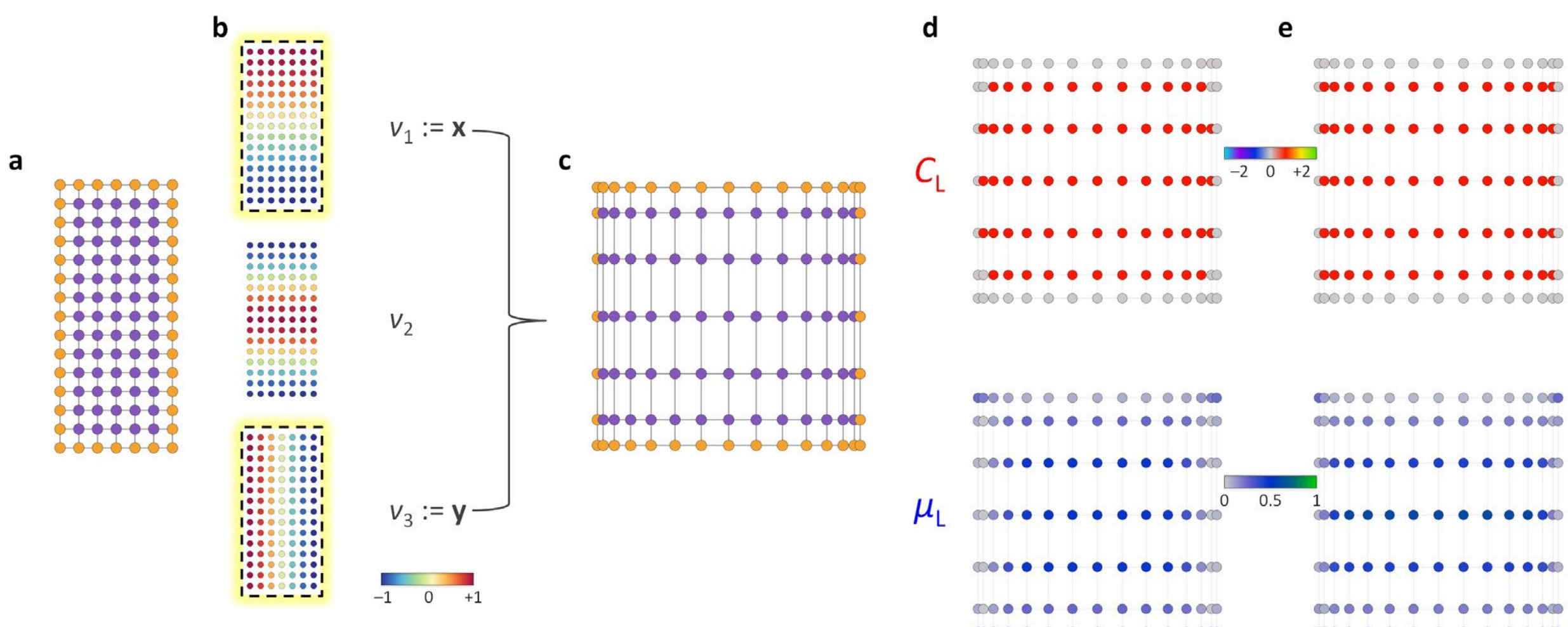

**Fig. S1. Embedding of a rectangular Hofstadter lattice. a,** Hofstadter lattice of 7 × 15 sites. **b**, The lowest nontrivial eigenstates $v_1$, $v_2$, … of $G$, where $v_1$ and $v_3$ define the site coordinates $x$ and $y$ and the corresponding position operators $X$ and $Y$. **c,d**, The Hamiltonian-derived embedding of the lattice (**c**) and the resulting $C_L$ and $\mu_L$ (**d**). **e**, $C_L$ and $\mu_L$ obtained with the original embedding in **a**, plotted on our Hamiltonian-derived embedding of **c**. All other parameters are identical to those in Fig. 2 of the main text.

**Supplementary Note S2. Topological diagnostics of the original Hofstadter lattice**

Figure S2 shows the spectral localizer evaluated on the original site positions of the 11 × 11 Hofstadter lattice, which serves as the reference for the Hamiltonian-derived embedding in Figs. 2d and 2e. The bulk index $C_L = +1$ and the finite local gap $\mu_L$ (Figs. S2a and S2b), as well as the butterfly of the bulk-averaged $\langle C_L \rangle$ over α and E (Fig. S2c), are reproduced on the derived embedding, where $C_L = +1$ is recovered on 85% of the bulk sites at the design point. The remaining sites lie near the corners of the embedding, where the local gap closes.

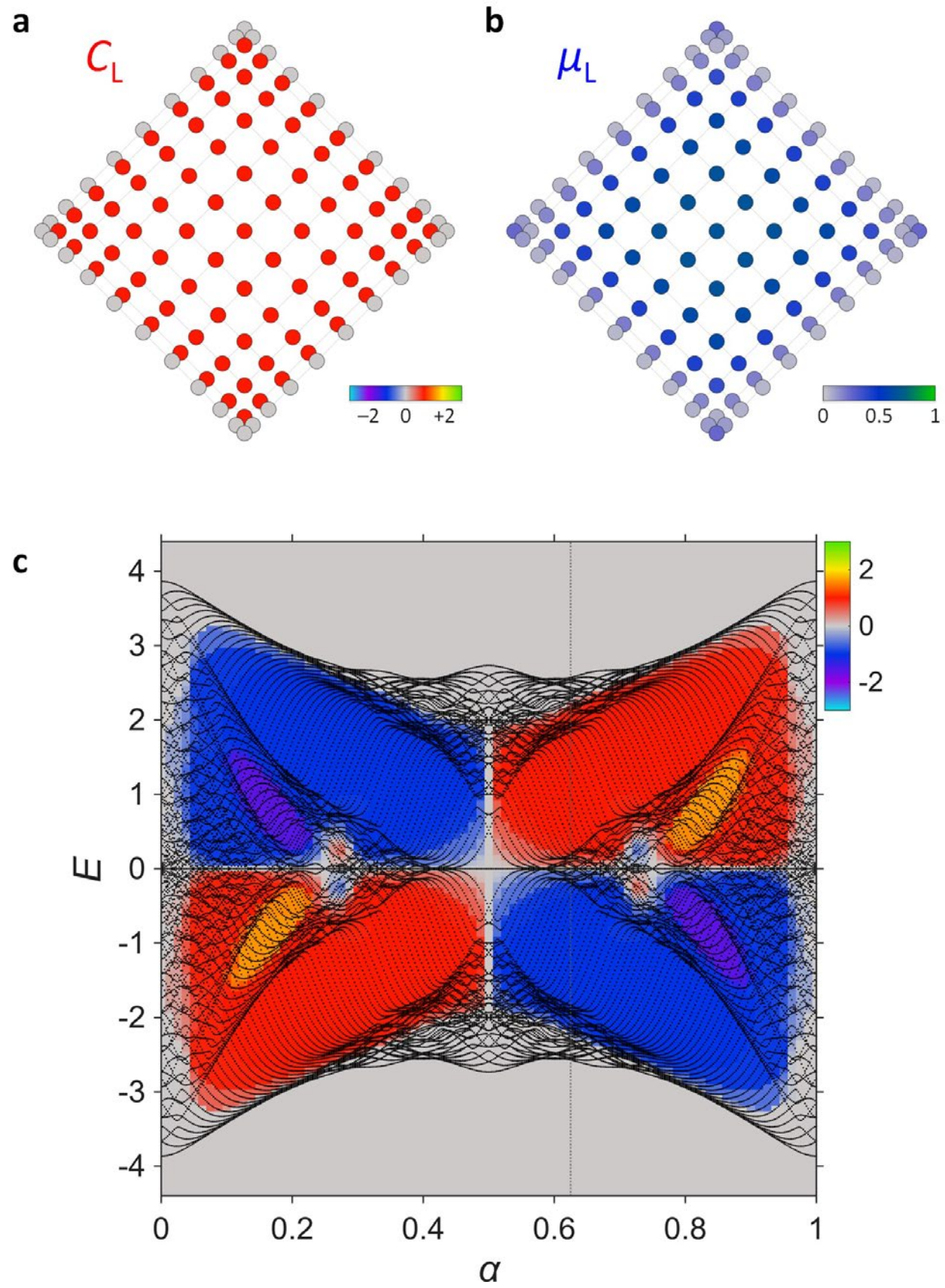

**Fig. S2. Topological diagnostics on the original embedding of the Hofstadter lattice. a,b**, $C_L$ (**a**) and $\mu_L$ (**b**). **c**, Bulk-averaged $\langle C_L \rangle$ as a function of 2π-normalized flux per plaquette $\alpha$ and energy $E$, with the eigenvalue spectrum (black dots). **a-c** correspond to Figs. 2d and 2e of the main text, with all other parameters identical.

## Supplementary Note S3. Topological diagnostics of various lattices

Figures S3 and S4 compare the Hamiltonian-derived embedding with the original embedding for the four lattices treated in Figs. 2f-2i of the main text. For every lattice, $C_L$ and $\mu_L$ on the Hamiltonian-derived embedding (upper maps in Fig. S3) agree with those on the original site positions (lower maps). The butterflies on the original embeddings (Fig. S4) also coincide with those on the Hamiltonian-derived embeddings (Figs. 2f-2i of the main text) over the full ranges of $\alpha$ and $\varphi$. Most of the differences between the two embeddings occur at the boundary between the bulk and edge regions, where the local gap is small in both embeddings. For the hyperbolic lattice, the local gap is small throughout, reflecting the fragility of topological phases observed in this non-Euclidean lattice [1].

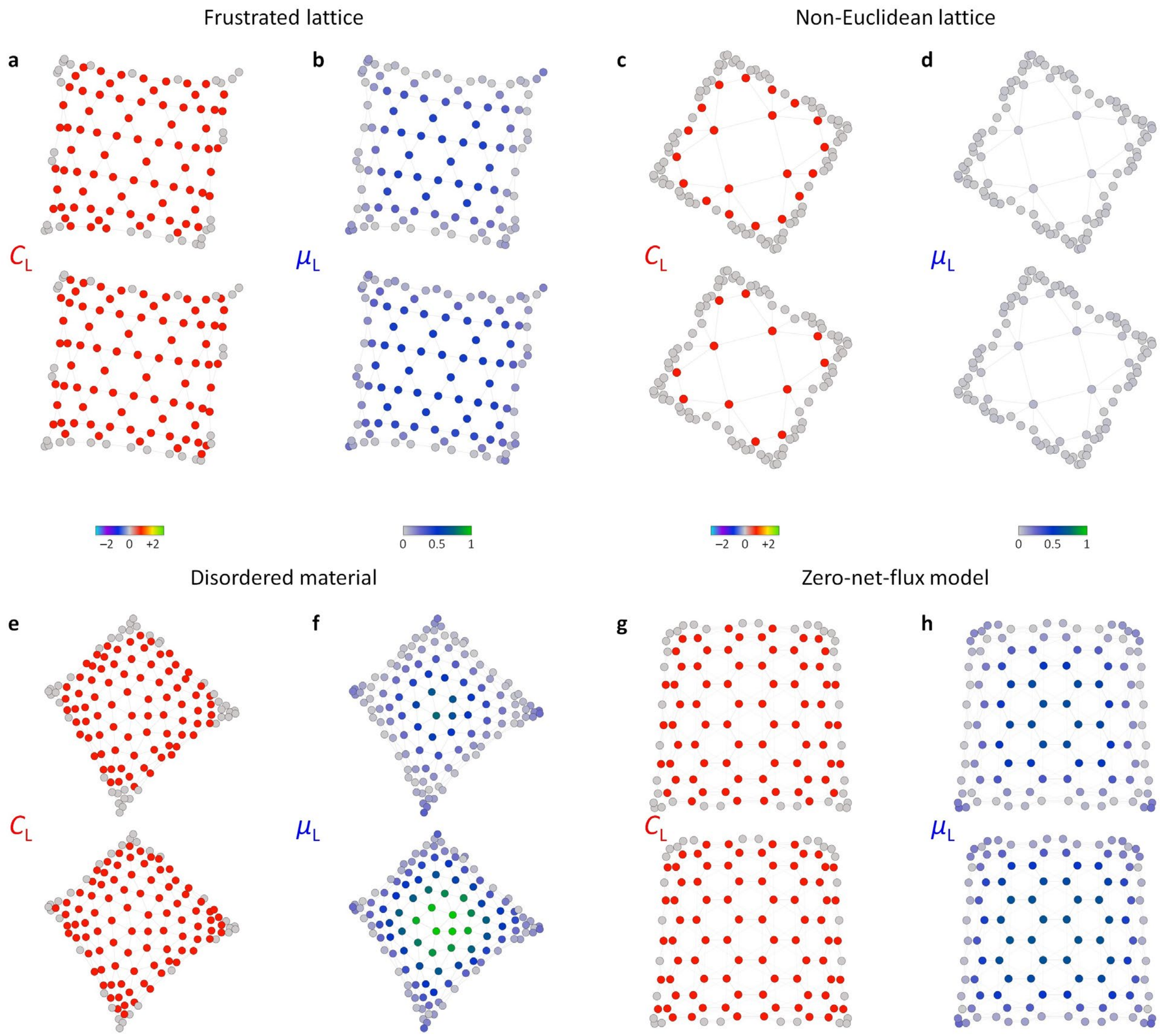

**Fig. S3. Topological diagnostics of various lattices.** The kagome lattice (**a,b**), the hyperbolic {4, 6} lattice (**c,d**), the SHU lattice (**e,f**), and the Haldane model (**g,h**): $C_L$ (**a,c,e,g**) and $\mu_L$ (**b,d,f,h**). In each panel, the upper map is evaluated on the Hamiltonian-derived embedding and the lower map on the original embedding, both plotted on the Hamiltonian-derived coordinates. Lattice parameters and design points are given in Methods.

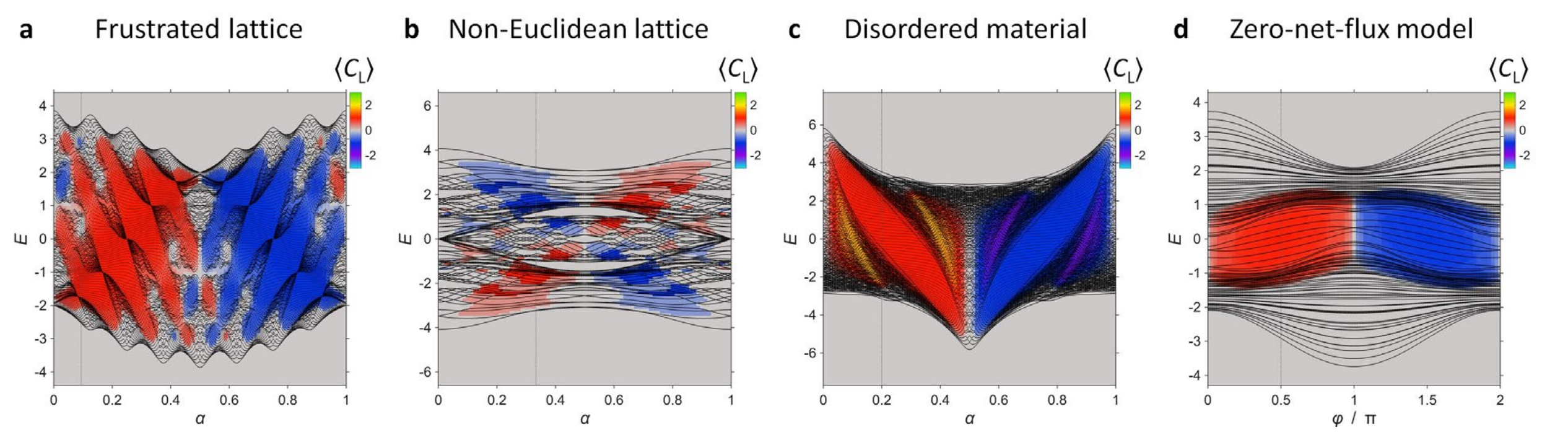

**Fig. S4. Hofstadter butterflies on the original embeddings of various lattices.** Bulk-averaged $\langle C_{\mathrm{L}} \rangle$ as a function of the 2π-normalized flux $\alpha$ and the energy $E$ for the kagome (**a**), hyperbolic (**b**), and SHU (**c**) lattices, and of the next-nearest-neighbour phase $\varphi$ for the Haldane model (**d**), evaluated with the original site positions. The eigenvalue spectrum (black dots) and the design values (vertical lines) are the same as in Figs. 2f-2i, which show the corresponding butterflies on the Hamiltonian-derived embeddings.

**Supplementary Note S4. Small-world regime of the scanned networks**

Figure S5 illustrates the small-world regime of the networks used in Fig. 3 of the main text. The path length drops steeply near $p = 0.01$, whereas the clustering survives up to $p \approx 0.1$. The networks with high clustering and short path length, $0.01 \lesssim p \lesssim 0.1$, are therefore classified into the small-world regime. The optimum of Fig. 3d of the main text lies at its upper end.

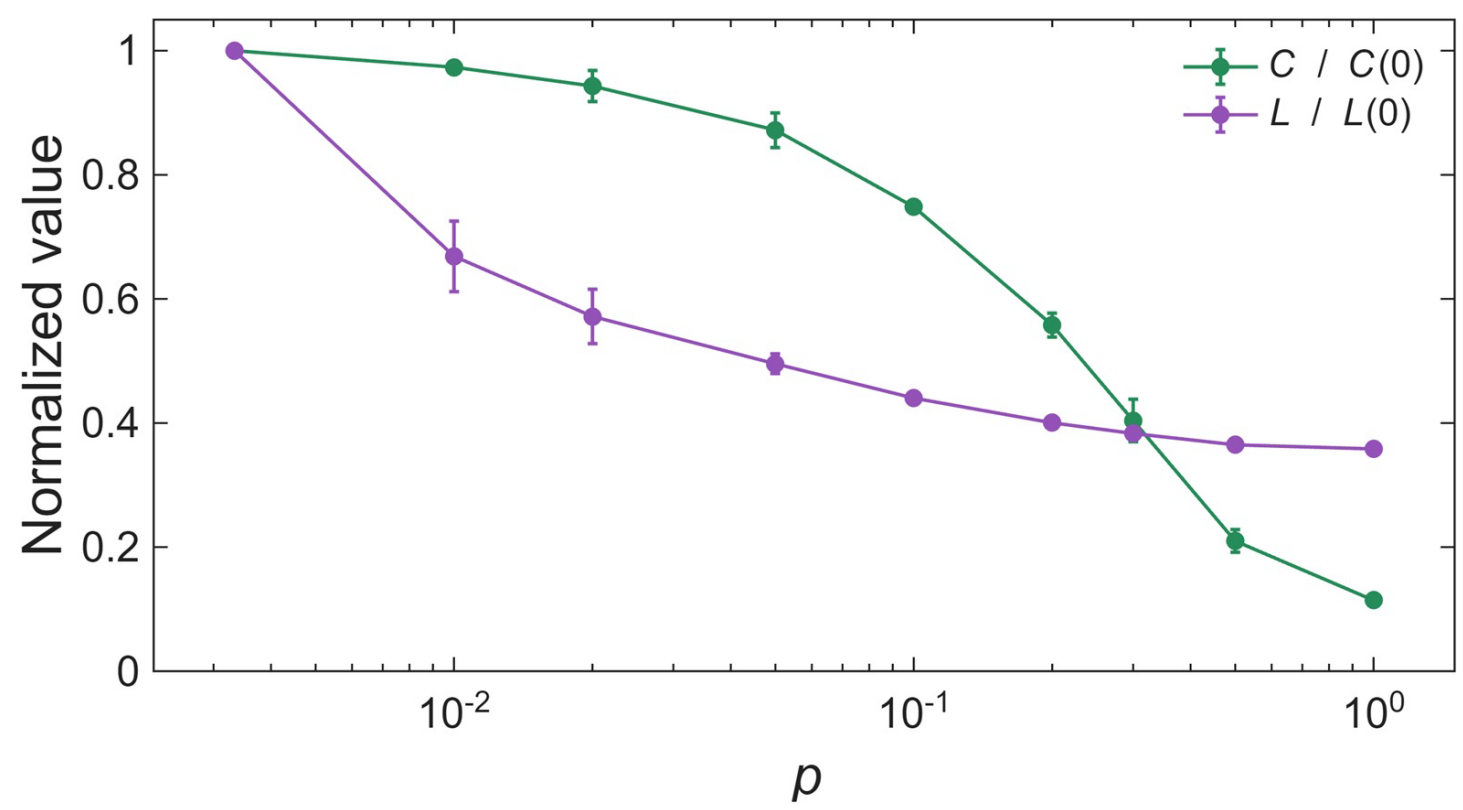

**Fig. S5. Small-world measures of the Watts-Strogatz networks.** Clustering coefficient $C$ and average path length $L$, normalized by their values at $p = 0$, as functions of the rewiring probability $p$ ($N = 120$, $k = 10$). Dots and error bars denote the mean and standard deviation over four network realizations per $p$. The point at the left edge denotes $p = 0$.

**Supplementary Note S5. Topological phases designed with position optimization**

Figure S6 applies the Hamiltonian-derived embedding to the small-world topological insulator designed in our previous work [2], whose link gauge fields and site positions were optimized jointly for $C_\mathrm{L} = +1$ on the inner nodes (Fig. 1e in the main text). On the optimized positions, the design is realized precisely: every bulk node has $C_\mathrm{L} = +1$ and every edge node has $C_\mathrm{L} = 0$, with a large bulk local gap $\mu_\mathrm{L}$ (Fig. S6e). However, the connectivity of this network, as characterized by the Hamiltonian-derived embedding (Fig. S6b), is that of a nearly regular ring, on which the bulk and edge nodes become intermixed (Fig. S6c). Owing to the absence of a well-defined bulk, no node acquires a nontrivial local Chern index ($C_\mathrm{L} \neq 0$) on the Hamiltonian-derived embedding (Fig. S6d), although the local gap remains finite. The designed topological phase therefore relies on the optimized positions themselves, which our connectivity-based embedding cannot reconstruct. This result illustrates that the Hamiltonian-derived embedding captures the topological phases inherited from local Hamiltonians but not every topological phase available in a complex network.

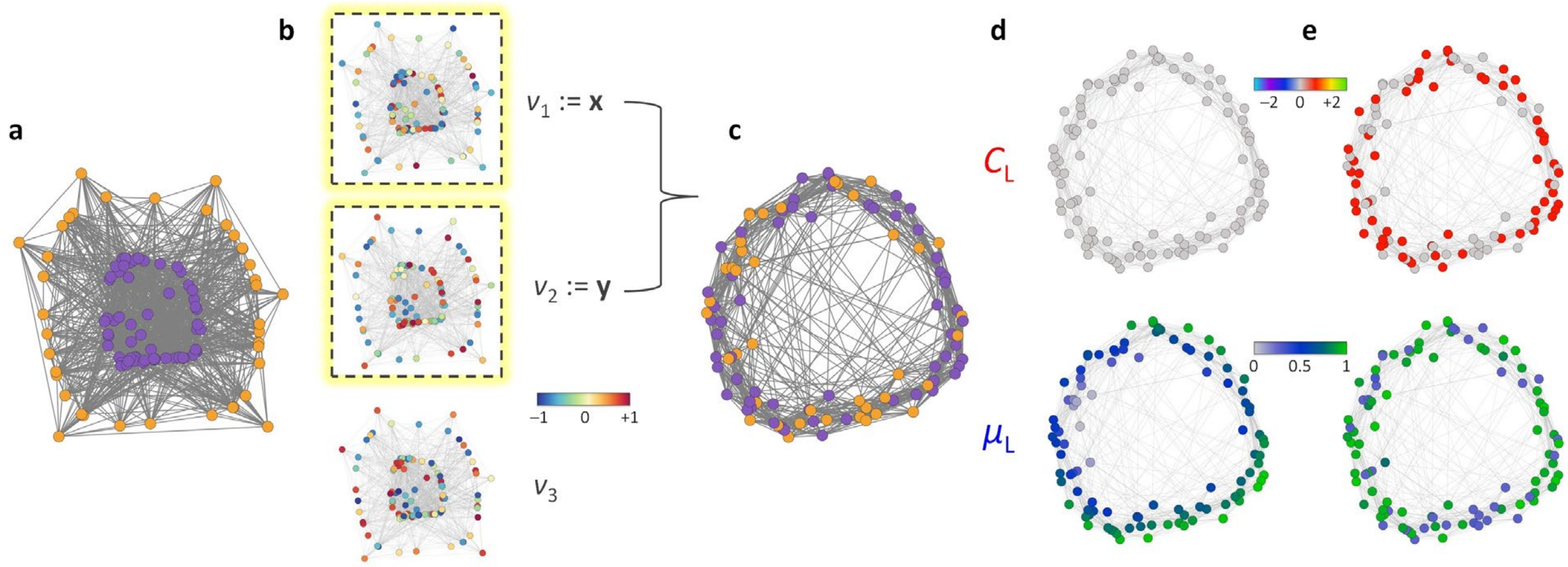

**Fig. S6. Hamiltonian-derived embedding of the position-optimized small-world topological insulator. a**, Embedding optimized jointly with the link gauge fields in [2]. Purple and amber sites denote the bulk and edge, respectively. **b,** The lowest nontrivial eigenstates $v_1$, $v_2$, … of $G$, where $v_1$ and $v_2$ define the site coordinates $x$ and $y$ and the corresponding position operators $X$ and $Y$. **c**,**d**, The Hamiltonian-derived embedding of the network (**c**) and the resulting $C_L$ and $\mu_L$ (**d**). **e**, $C_L$ and $\mu_L$ obtained with the original embedding in **a**, plotted on our Hamiltonian-derived embedding of **c**.

## Supplementary References

[1] S. Yu, X. Piao, and N. Park, Topological hyperbolic lattices, Phys. Rev. Lett. **125**, 053901 (2020).

[2] S. Yu, X. Piao, and N. Park, Topological insulators on complex networks, arXiv preprint arXiv:2610.02888 (2026).